\documentclass{PoS}

\title{Morphological properties of blazar-induced gamma-ray haloes}

\ShortTitle{Morphological properties of blazar-induced gamma-ray haloes}

\author{\speaker{Rafael~{Alves Batista}}\\
        University of Oxford, Department of Physics - Astrophysics \\
        DWB, 1 Keble Road, Oxford OX1~3RH, United Kingdom\\
        E-mail: \email{rafael.alvesbatista@physics.ox.ac.uk}}

\author{Andrey Saveliev\\
Immanuel Kant Baltic Federal University, Institute of Physics, Mathematics and Information Technology, 236041 Kaliningrad, Russia \\
        E-mail: \email{andrey.saveliev@desy.de}}

\abstract{
At TeV energies and above gamma rays can induce electromagnetic cascades, whose charged component is sensitive to intervening intergalactic magnetic fields (IGMFs). When interpreting gamma-ray measurements in the energy range between a few GeV and hundreds of TeV, one has to carefully account for effects due to IGMFs, which depend on their strength and power spectrum. Therefore, gamma-ray-induced electromagnetic cascades can be used as probes of cosmic magnetism, since their arrival distribution as well as spectral and temporal properties can provide unique information about IGMFs, whose origin and properties are currently poorly understood. In this contribution we present an efficient three-dimensional Monte Carlo code for simulations of gamma-ray propagation. We focus on the effects of different configurations of IGMFs, in particular magnetic helicity and the power spectrum of stochastic fields, on the morphology of the arrival directions of gamma rays, and discuss the prospects for detecting pair haloes around distant blazars.
}

\FullConference{35th International Cosmic Ray Conference --- ICRC2017\\
		10--20 July, 2017\\
		Bexco, Busan, Korea}

\begin{document}

\section{Introduction}

The origin of magnetic fields in the universe is a long-standing problem in cosmology. While there are a few astrophysical mechanisms within sources capable of generating the currently observed magnetic fields, many models postulate a cosmological origin via, for example, electroweak and QCD phase transitions, leptogenesis, and inflation, among others. For reviews the reader can refer to e.g.~\cite{kandus2011a}. Observations of very-high energy gamma rays using image air Cherenkov telescopes (IACTs) can be used to constrain properties of intergalactic magnetic fields (IGMFs).

Very-high-energy (VHE) gamma rays, whose energies exceed $\sim 100 \; \mathrm{GeV}$, may interact with diffuse extragalactic background radiation fields, namely the extragalactic background light (EBL) and the cosmic microwave background (CMB), generating electron-positron pairs via pair production ($\gamma + \gamma_{bg} \rightarrow e^+ + e^-$, wherein the subscript `bg' refers to the CMB and EBL). The electrons and positrons\footnote{Henceforth we will refer to electrons and positrons collectively as ``electrons''.} up-scatter photons contained in the CMB and EBL via inverse Compton scattering ($e^\pm + \gamma_{bg} \rightarrow e^\pm + \gamma$). As a consequence, the flux of VHE gamma rays from distant sources is attenuated due to pair production, exponentially decreasing as $\exp(\tau_{\gamma\gamma})$, where $\tau_{\gamma\gamma}$  is the pair production optical depth~\cite{aharonian1994a,plaga1994a}. 

There are evidences supporting deviations from the expected opacity at very-high energies~\cite{albert2008a,horns2012a,dominguez2013a}. Some of the solutions proposed include: ultra-high-energy cosmic rays initating the cascades~\cite{essey2010a,essey2011a}; strong intergalactic magnetic fields (IGMFs)~\cite{neronov2010a,taylor2011a}; oscillation of gamma rays into axion-like particles~\cite{deangelis2007a,mirizzi2009a}. These processes affect the inferred energy spectrum of gamma rays.

The extension of blazar-induced pair haloes has been studied in many works (see e.g. Refs.~\cite{dolag2009a,eungwanichayapant2009a}); their morphology, however, has been addressed in only a few studies~\cite{neronov2010b,long2015a,alvesbatista2016b,duplessis2017a,fitoussi2017a}, particularly those related to the helicity of IGMFs.

The topology of the magnetic field is related to a quantity known as magnetic helicity, defined as $\mathcal{H} = \int d^3 x \, \vec{B} \cdot \vec{A}$, with $\vec{B} \equiv \vec{\nabla} \times \vec{A}$. Evidences have been found supporting the existence of helical IGMFs using data from Fermi~\cite{tashiro2014a,chen2015a}. The search for signatures of helical fields is important because they may relate to the baryonic asymmetry of the universe through CP violation~\cite{vachaspati2001a}. Moreover, the handedness of helical fields are indicators of the exact magnetogenic process.

\section{Code}

There are some notorious works on the propagation of VHE gamma-ray-induced electromagnetic cascades using Monte Carlo methods. Most of them rely on the small angle approximation; the well-known code ELMAG~\cite{kachelriess2012a} is an example thereof. There are a few works making use of three-dimensional simulations~\cite{neronov2010b,alvesbatista2016b,fitoussi2017a}, which allows detailed studies of the morphology of pair haloes around distant blazars.

In this work we provide a full three-dimensional Monte Carlo treatment of the development of gamma-ray-induced cascades. The code is based on the modular structure of CRPropa3~\cite{alvesbatista2016a} and has been successfully used in previous works~\cite{alvesbatista2016b,alvesbatista2017a}. It takes into account all relevant interaction and energy-loss processes, as well as deflections of the charged component of the cascade in intervening IGMFs.

The deflection of a gamma ray with initial energy $E_i$ located at a distance $D$ from Earth, corresponding to redshift $z$, due to a magnetic field with rms strength $B$, is given by~\cite{neronov2009a}:
\begin{equation}
	\delta(E) = 0.05^\circ \kappa \frac{1}{(1 + z)^4} \left( \frac{B}{10^{-15} \; \mathrm{G}} \right)  \left( \frac{E_i}{10^{11} \; \mathrm{eV}} \right)^{-1} \left( \frac{D}{1 \; \mathrm{Gpc}} \right)^{-1}  \left( \frac{E}{10^{13} \; \mathrm{eV}} \right)^{-1},
	\label{eq:defl}
\end{equation}
where $E$ is the energy at which the gamma ray is observed at Earth, and $\kappa$ is a scaling factor to account for intrinsic uncertainties of EBL models.

\section{Impact of the magnetic field strength, power spectrum, and coherence length}

For the examples presented in this section we take the blazar 1ES0229+200, located at a redshift $z \approx 0.14$. This object has been studied across a wide range of wavelengths. We analyse the following observables: size of the pair halo and spectrum. These depend on properties of the intervening magnetic fields, namely its power spectrum, strength, and coherence length. In the following subsections we study the effects of these three properties on gamma-ray observables. 

In our simulation setup the magnetic field is comprised of periodically repeating uniform cubic subgrids of side 50 Mpc and resolution of approximately 50 kpc. The Kolmogorov power spectrum is characterised by the following proportionality relation: $\langle \vec{B}(\vec{k}) \cdot \vec{B}(\vec{k'}) \rangle \propto k^{-11/3}$; for the Batchelor spectrum $\langle \vec{B}(\vec{k}) \cdot \vec{B}(\vec{k'}) \rangle \propto k^{2}$. We assume further that the jet with half-opening angle of $5^\circ$ is pointing towards the observer ($\Theta = 0^\circ$). We have not considered the case of misaligned jets in this work. The magnetic field strengths considered are $B=10^{-15} \; \mathrm{G}$, $B=10^{-16} \; \mathrm{G}$, and $B=10^{-17} \; \mathrm{G}$, with coherence lengths $L_c = 0.03 \; \mathrm{Mpc}$, $L_c = 1 \; \mathrm{Mpc}$, and $L_c = 40 \; \mathrm{Mpc}$. We adopt the EBL model by Gilmore et al.~\cite{gilmore2012a}. 

\begin{figure}
  \centering
  \includegraphics[width=0.5\columnwidth]{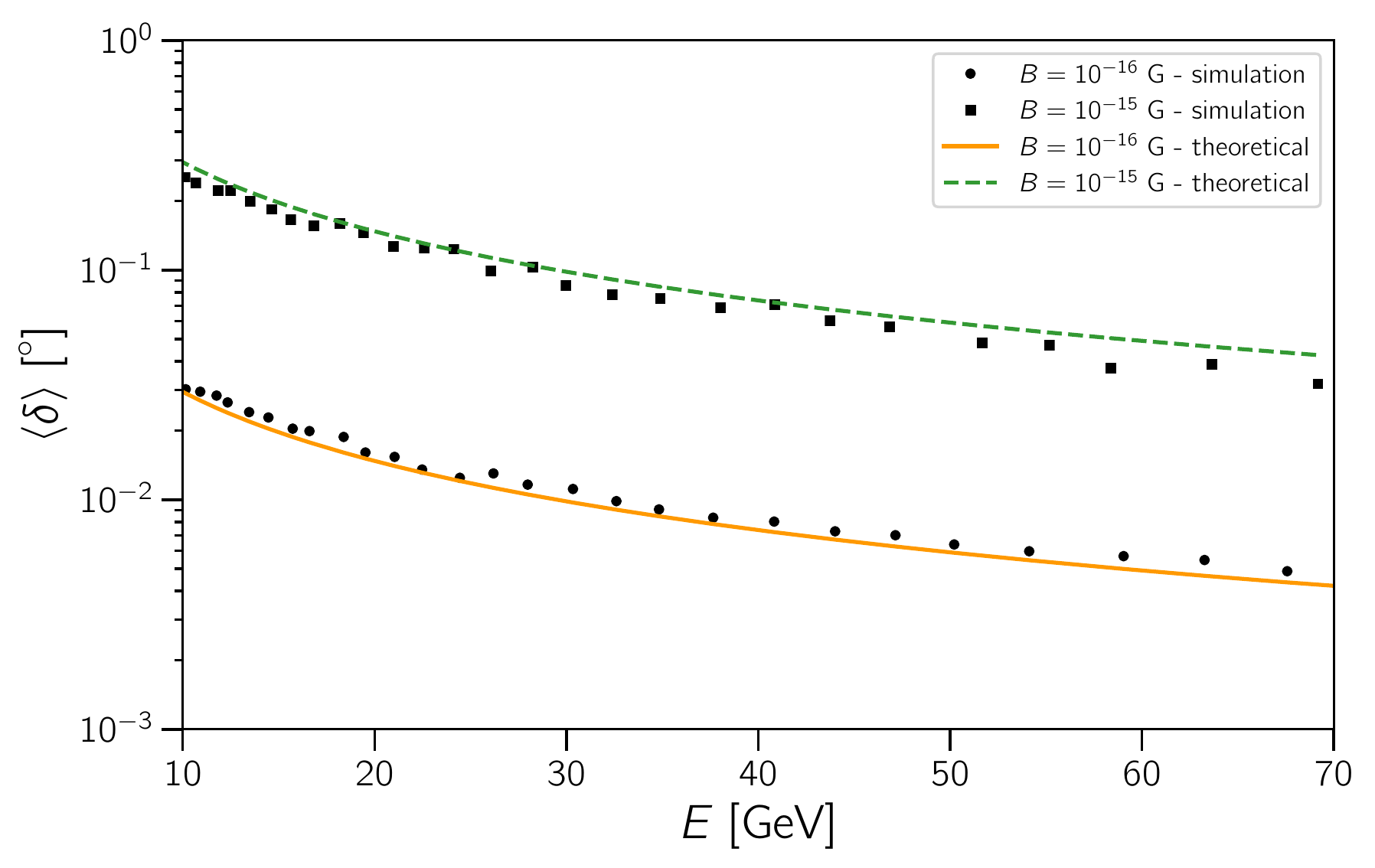}
  \caption{Comparison of simulation results (markers) with theoretical expectations from Eq.~\ref{eq:defl}~\cite{neronov2009a} for a source located at $z \approx 0.14$ emitting gamma rays with energies $E_i = 10^{14} \; \mathrm{eV}$. The magnetic field is assumed to have a Kolmogorov spectrum with coherence length $L_c \approx 1 \; \mathrm{Mpc}$; we assume that $\kappa \approx 0.75$.}
  \label{fig:check}
\end{figure}

In order to test the code, we compare theoretical estimates from Eq.~\ref{eq:defl} with our simulations. We consider a scenario in which a source at $z \approx 0.14$ emits gamma rays of energy $E_i = 10 \; \mathrm{TeV}$ within a tightly collimated jet. The magnetic field is assumed to be stochastic with a Kolmogorov spectrum and coherence length $L_c \simeq 1 \; \mathrm{Mpc}$. This is shown in Fig.~\ref{fig:check}.

The setup of our simulation is the following: the source is assumed to lie at the centre of a sphere with radius equal to its distance to Earth. Because we are interested in the effects of magnetic fields on the development of the electromagnetic cascades, unless otherwise stated, we obtain the arrival directions from the simulated data sets \emph{in the simulation frame}, as opposed to the observer frame. We have not properly estimated the arrival directions of the halo expected at Earth because our focus lies in the investigation of the effects of magnetic field properties. Consequently, {\emph Eq.~\ref{eq:defl} is not applicable} to the results shown in this section, having been used only for the checks described in Fig.~\ref{fig:check}. 
% Therefore, haloes measured on Earth do not resemble the ones shown in Figs.~\ref{fig:haloes}.

\begin{figure}
  \centering
  \includegraphics[width=0.50\columnwidth]{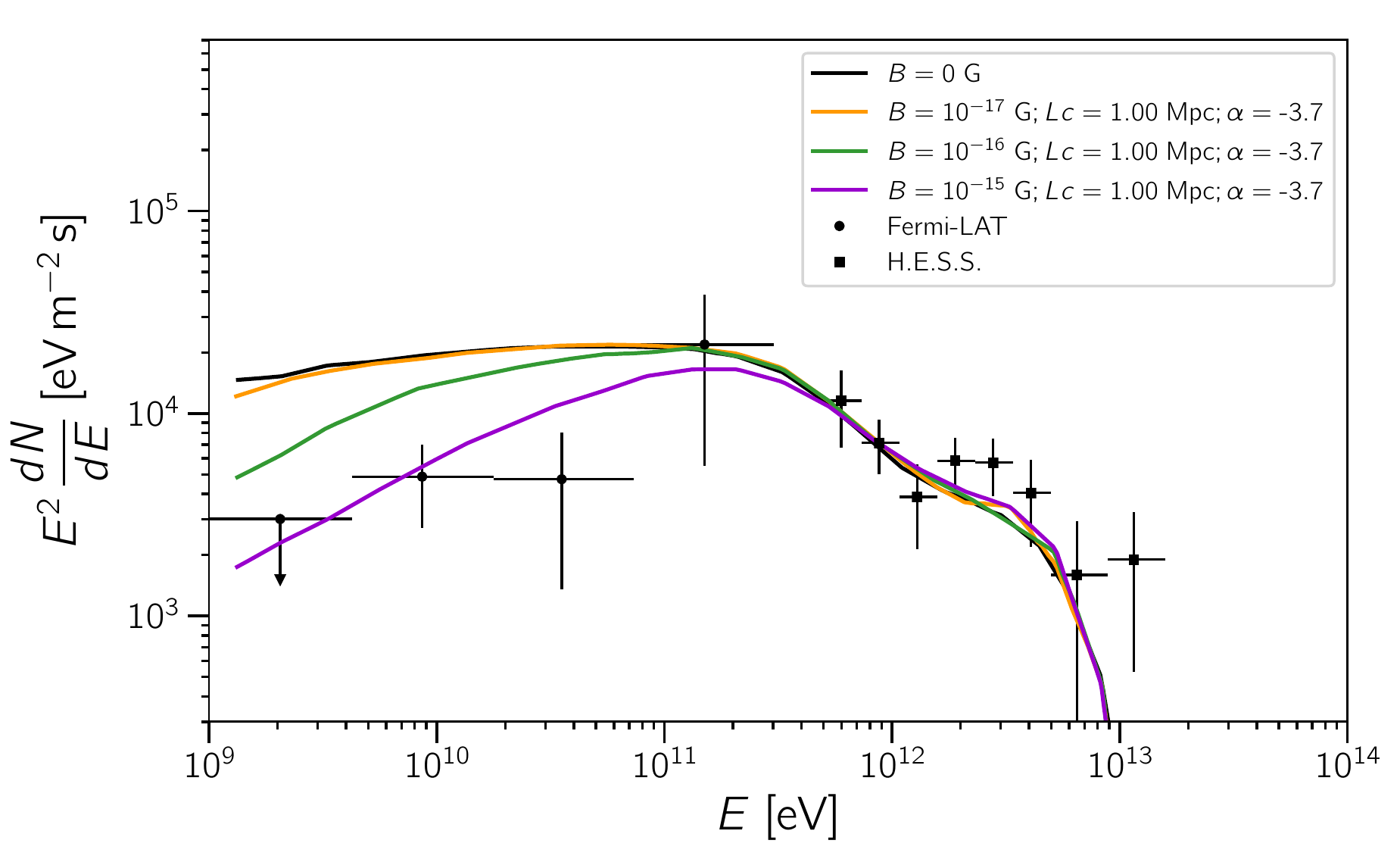}
  \caption{Energy spectrum of arriving particles from 1ES 0229+200. This object is located at $z \approx 0.14$, and is assumed to be emitting gamma rays with intrinsic spectrum $E^{-1.5}$ and cutoff energy Emax = 5 TeV, within a tightly collimated jet. The magnetic field has coherence length $L_c \approx 5 \mathrm{Mpc}$ and strengths $10^{-15} \; \mathrm{G}$ (violet), $10^{-16} \; \mathrm{G}$ (orange), $10^{-17} \; \mathrm{G}$ (green), and 0 (black). Measurements by H.E.S.S. (squares) and Fermi (circles) are shown for comparison. }
  \label{fig:specB}
\end{figure}

The simulations here presented are four-dimensional (3D + time). We impose a strict cut to them: only events arriving within a redshift window $\Delta z = 0.0005$ are considered.
We investigate the impact of different magnetic field strengths on the gamma-ray spectrum. This is shown in Fig.~\ref{fig:specB}.

\begin{figure}
  \centering
  \includegraphics[width=.495\textwidth]{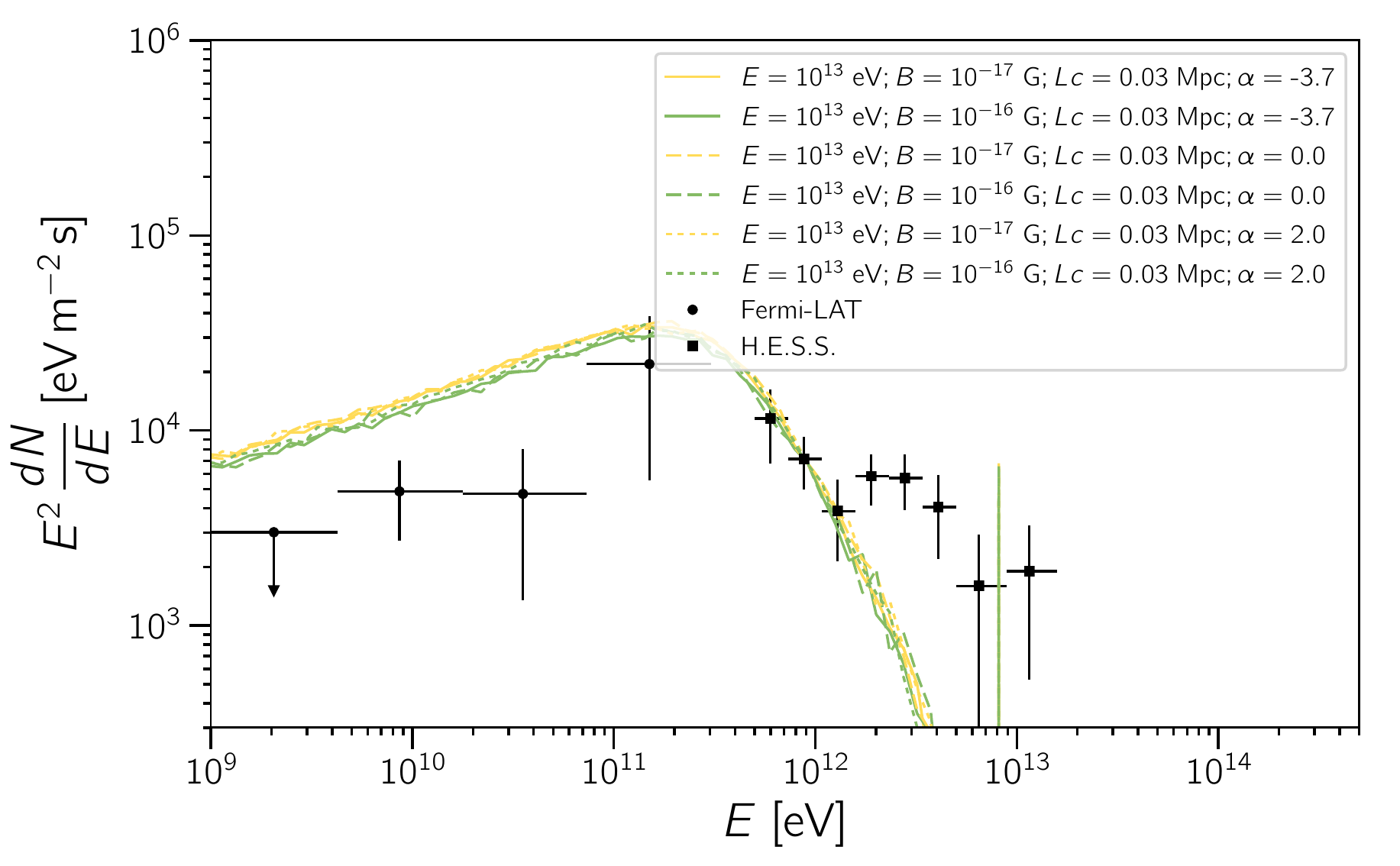}
  \includegraphics[width=.495\textwidth]{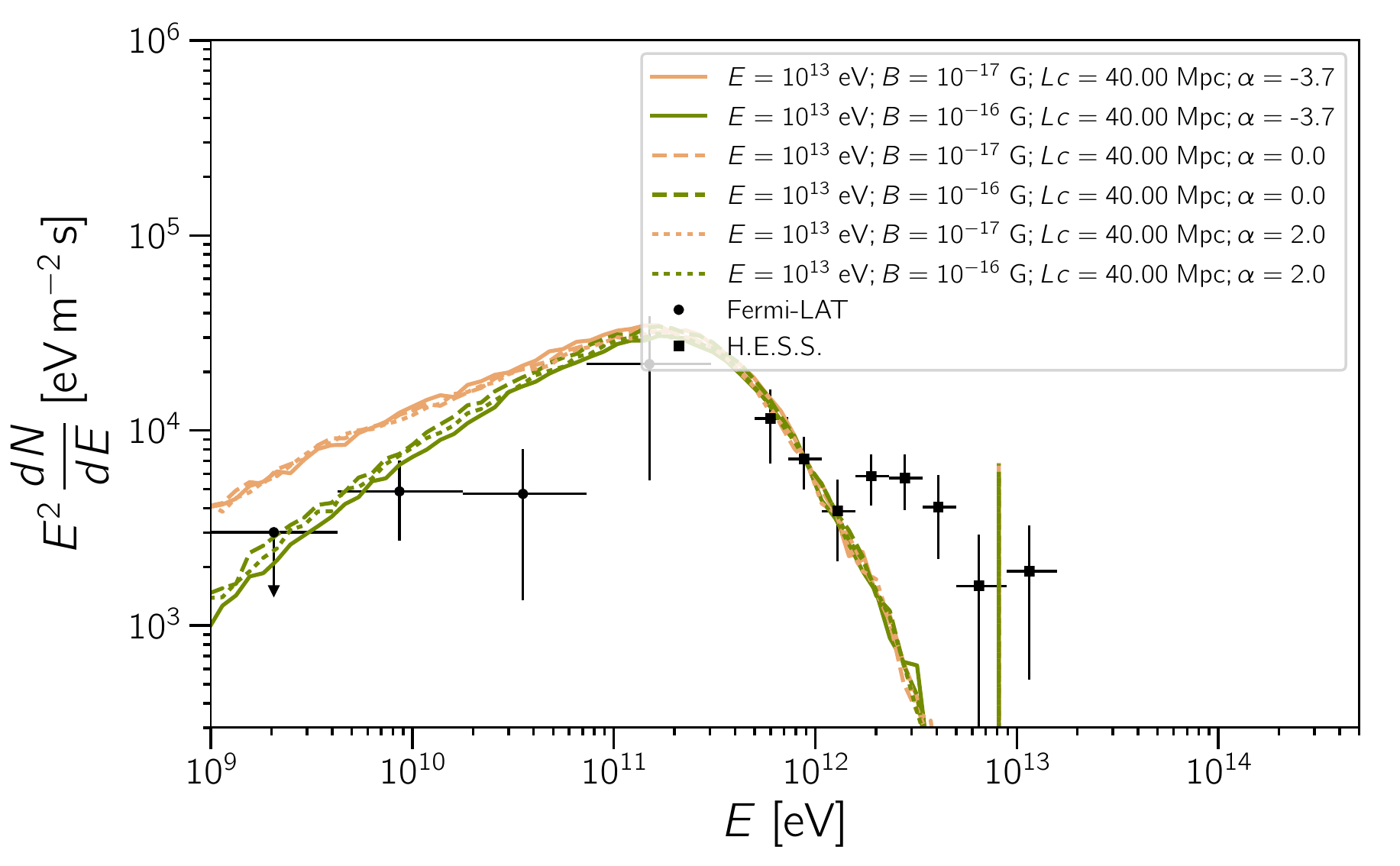}
  \includegraphics[width=.495\textwidth]{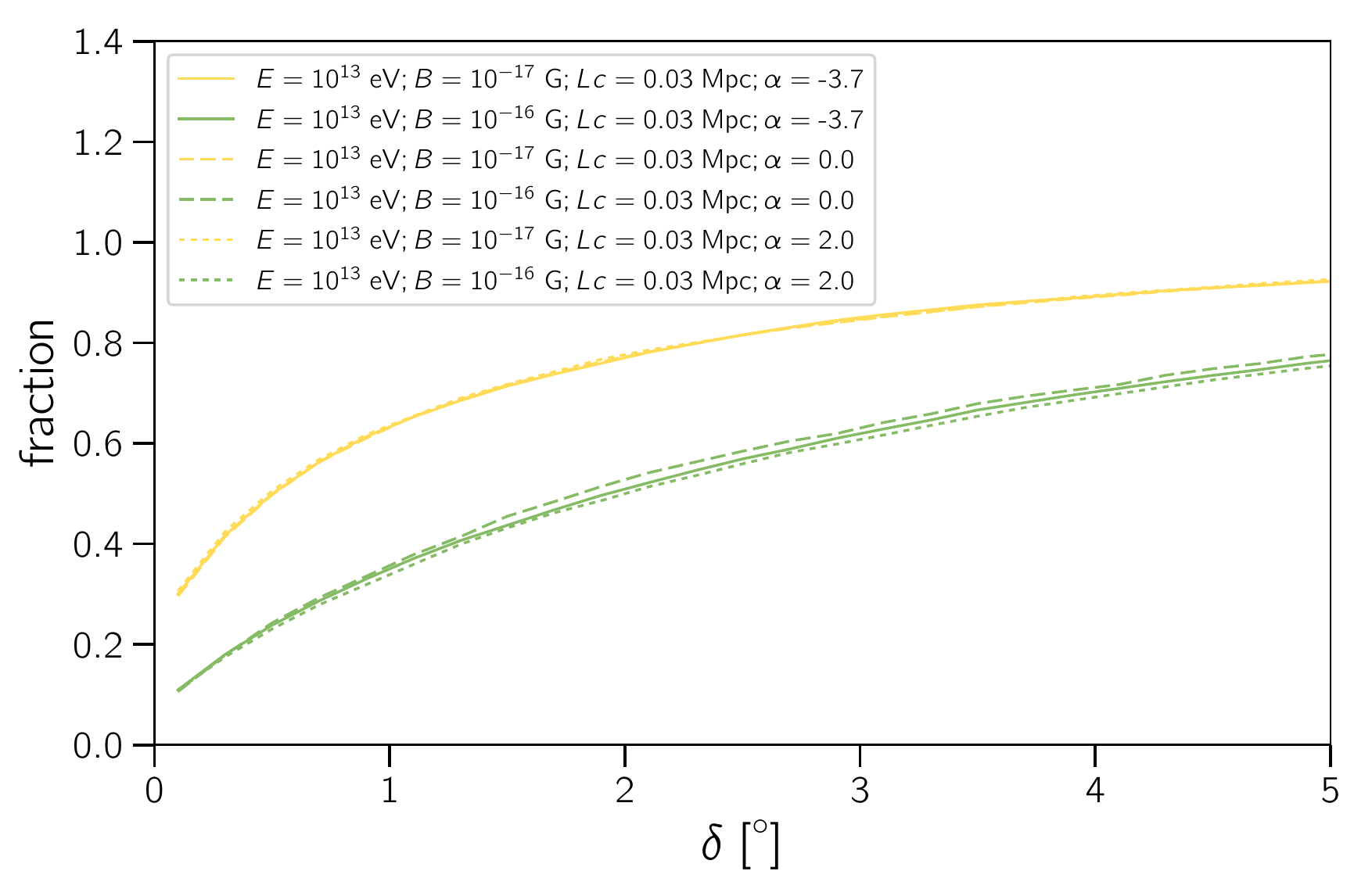}
  \includegraphics[width=.495\textwidth]{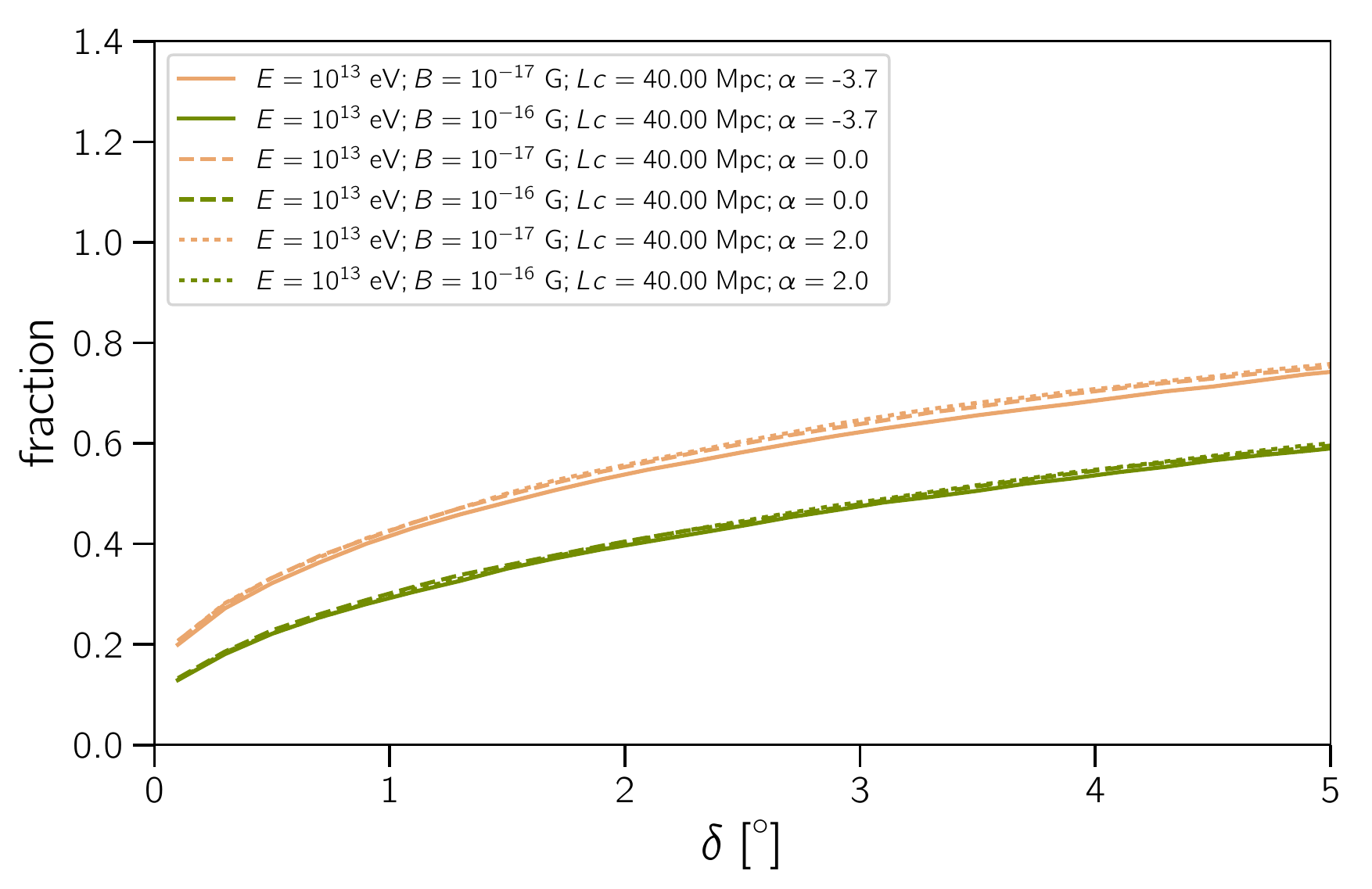}
  \caption{Spectra (top row) for a monochromatic source at $z \approx 0.14$ injecting gamma rays of energy 10 TeV. The fraction  of events contained within a radius $\delta$ is shown in the lower panels. Results are shown for $L_c = 30 \; \mathrm{kpc}$ (left), and for $L_c = 40 \; \mathrm{kpc}$ (right panels). }
  \label{fig:PS}
 \end{figure}

Poorly studied in the literature are the effects of different magnetic power spectra on the gamma-ray observables. To this end, we consider three scenarios for the power spectrum ($|\vec{B}|^2 \propto k^{\alpha}$), namely a Kolmogorov spectrum ($\alpha = -11/3$), white noise ($\alpha = 0$), and Batchelor spectrum ($\alpha = 2$). Note that these choices encompass a wide range of commonly found in the literature, including scale invariance from inflationary magnetogenesis ($\alpha = -3$). The containment radii (in the simulation's frame) is shown in the lower panels of Fig.~\ref{fig:PS} for two coherence lengths, namely $L_c = 30 \; \mathrm{kpc}$ and $L_c = 40 \; \mathrm{Mpc}$, together with the corresponding spectra (upper row).

We notice in Fig.~\ref{fig:PS} that the fraction of events contained within a given angular radius is inversely proportional to the magnetic field strength. Furthermore, the impact of $\alpha$ on this observable is virtually negligible. The same is also true for the spectra shown in Fig.~\ref{fig:PS}.

\begin{figure}[htb]
  \centering
  \includegraphics[width=0.32\columnwidth]{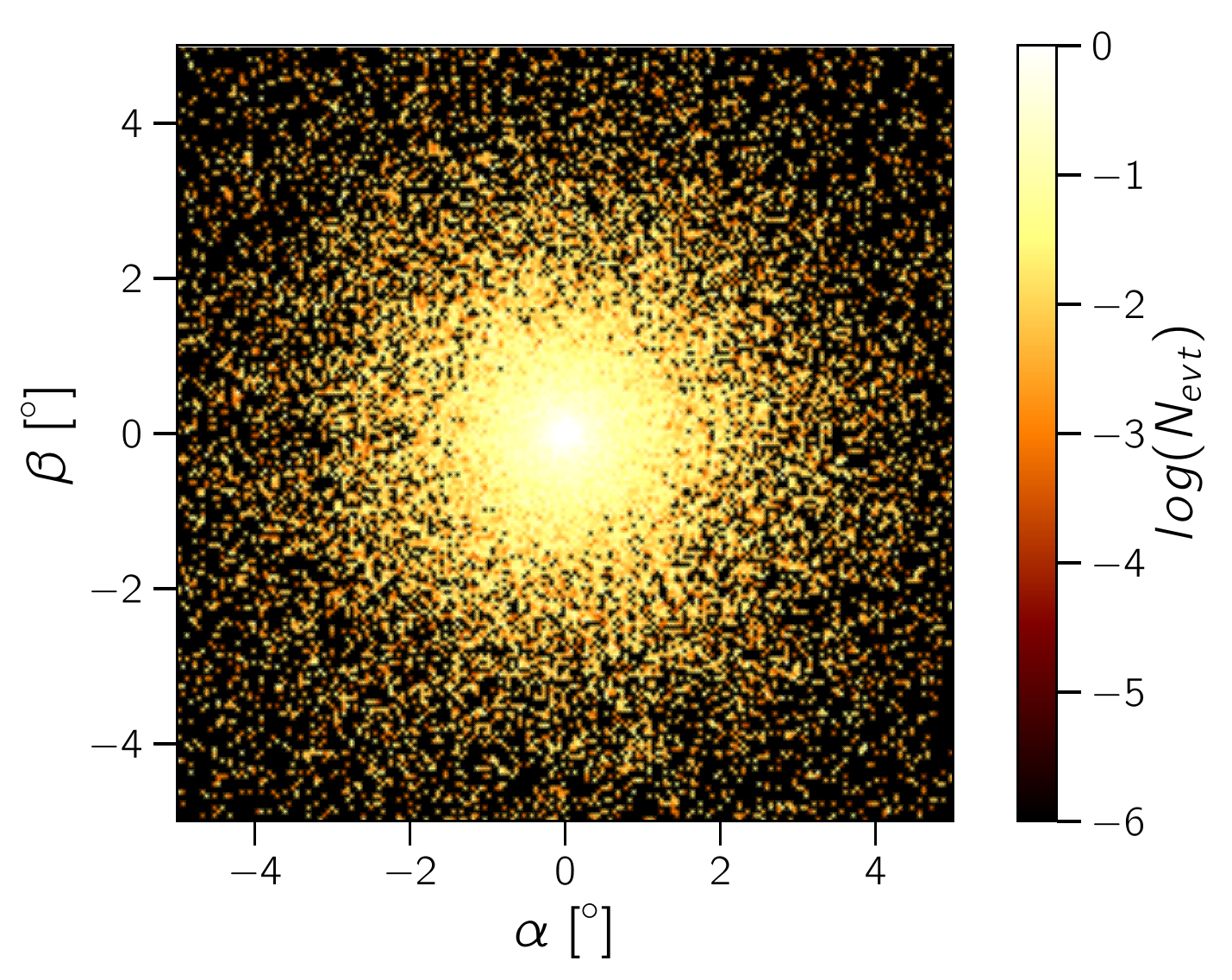}
  \includegraphics[width=0.32\columnwidth]{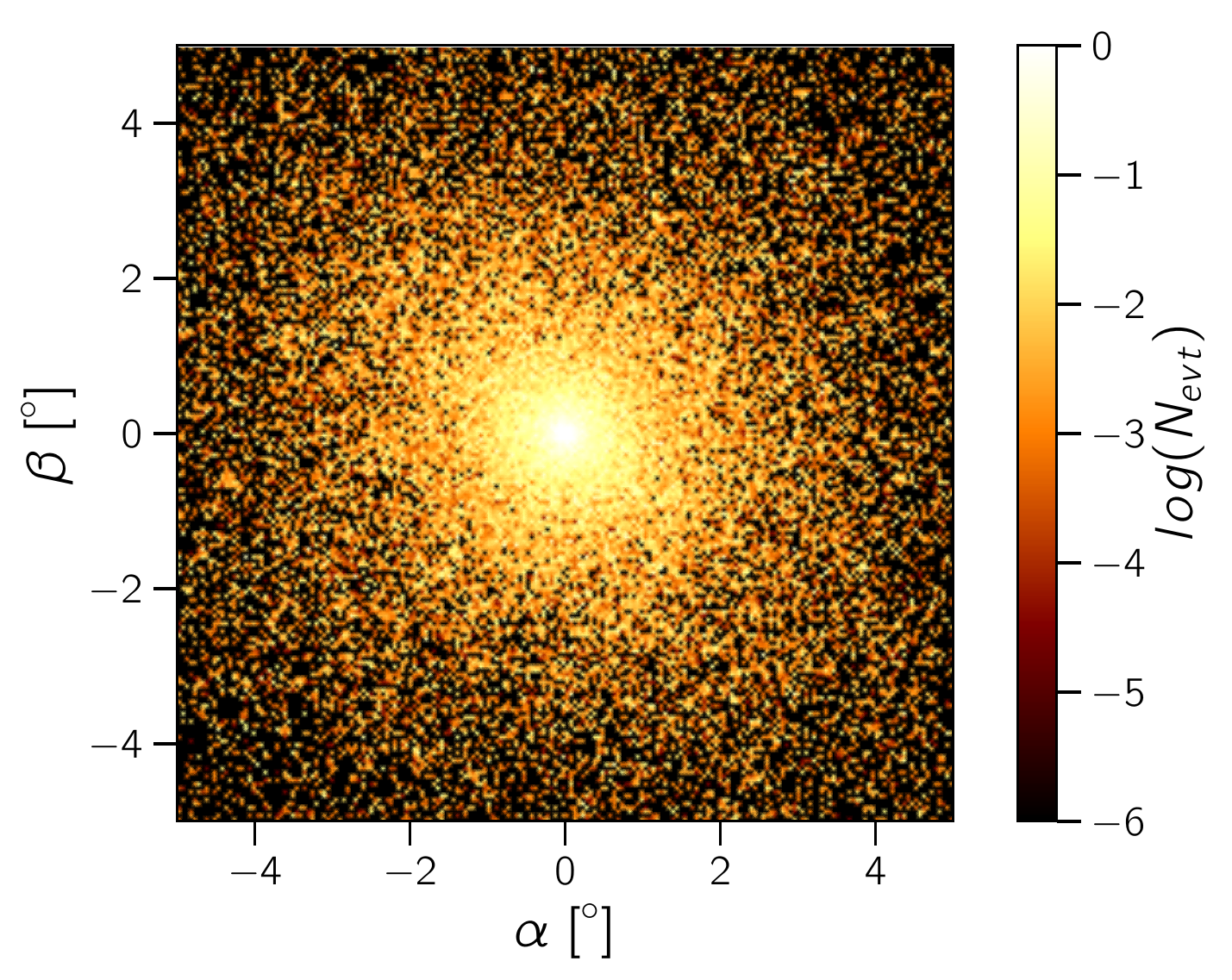}
  \includegraphics[width=0.32\columnwidth]{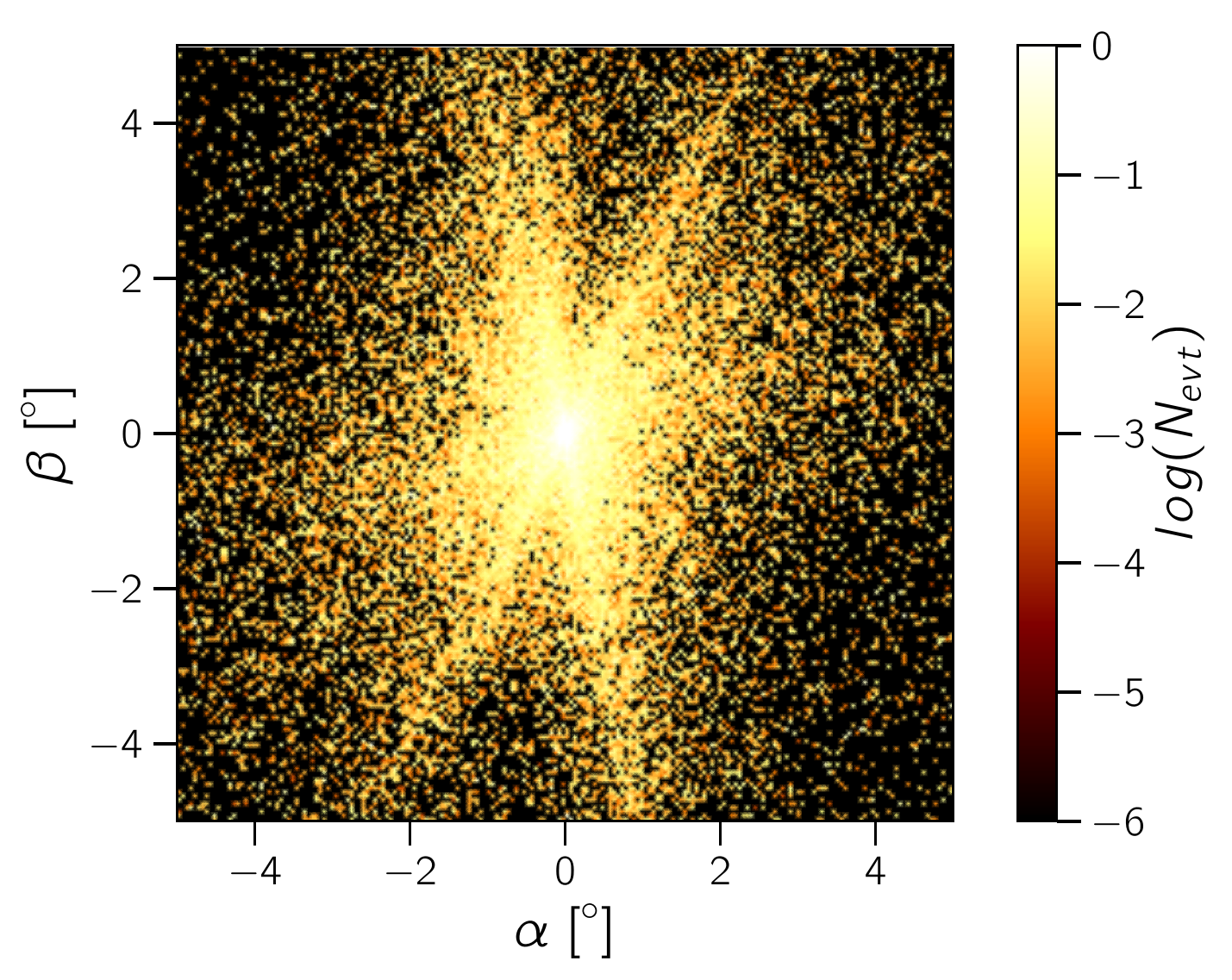}
  \caption{Haloes in the simulation's frame assuming a source at $z \approx 0.14$ emitting 10 TeV gamma rays. The magnetic field is assumed to have a spectral index $\alpha=0$, and strength $B=10^{-17} \; \mathrm{G}$, with coherence lengths $L_c = 30 \; \mathrm{kpc}$ (left), $L_c = 1 \; \mathrm{Mpc}$ (middle), and $L_c = 40 \; \mathrm{Mpc}$ (right panel).}
  \label{fig:haloes}
\end{figure}

\begin{figure}[htb]
  \centering
  \includegraphics[width=0.32\columnwidth]{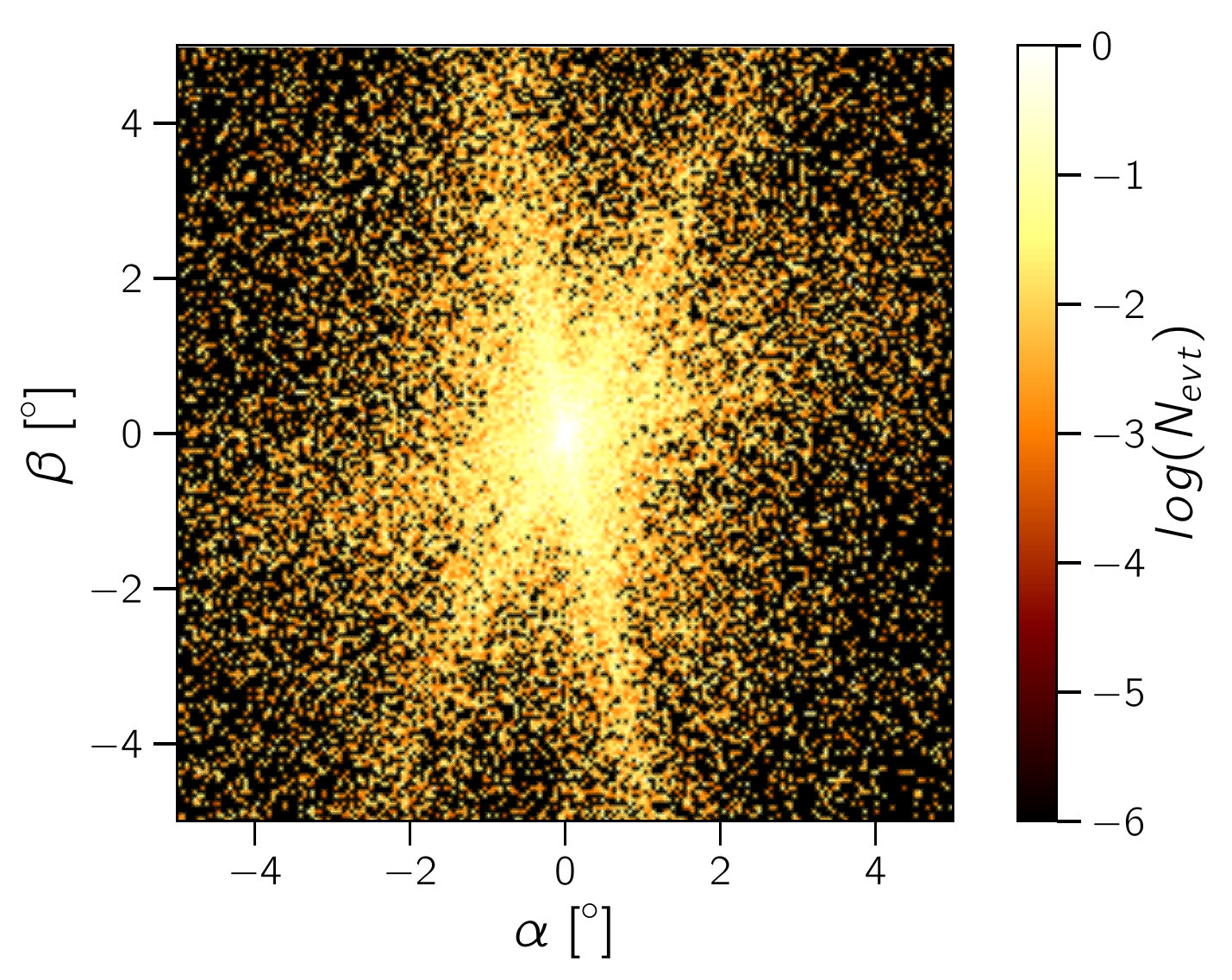}
  \includegraphics[width=0.32\columnwidth]{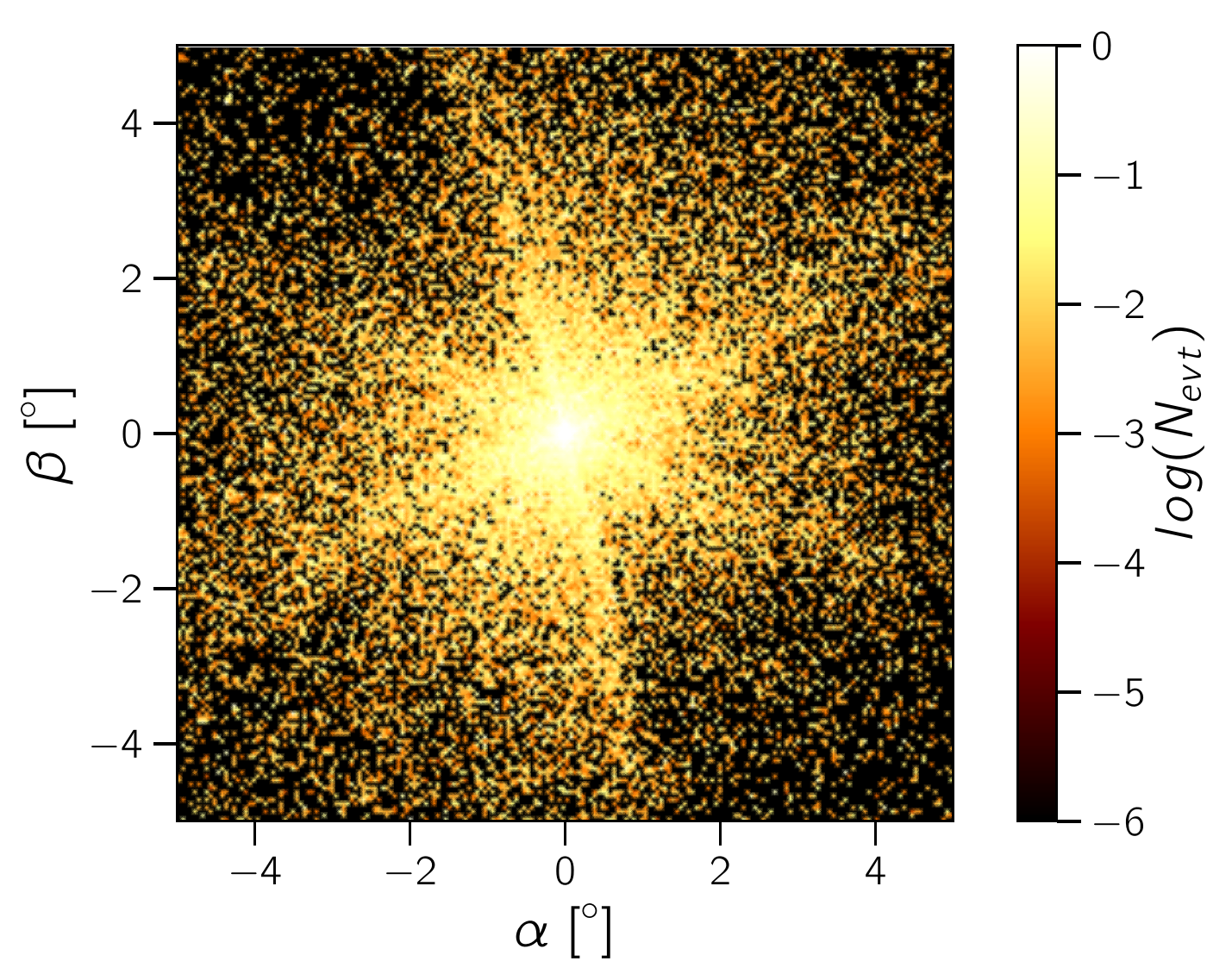}
  \includegraphics[width=0.32\columnwidth]{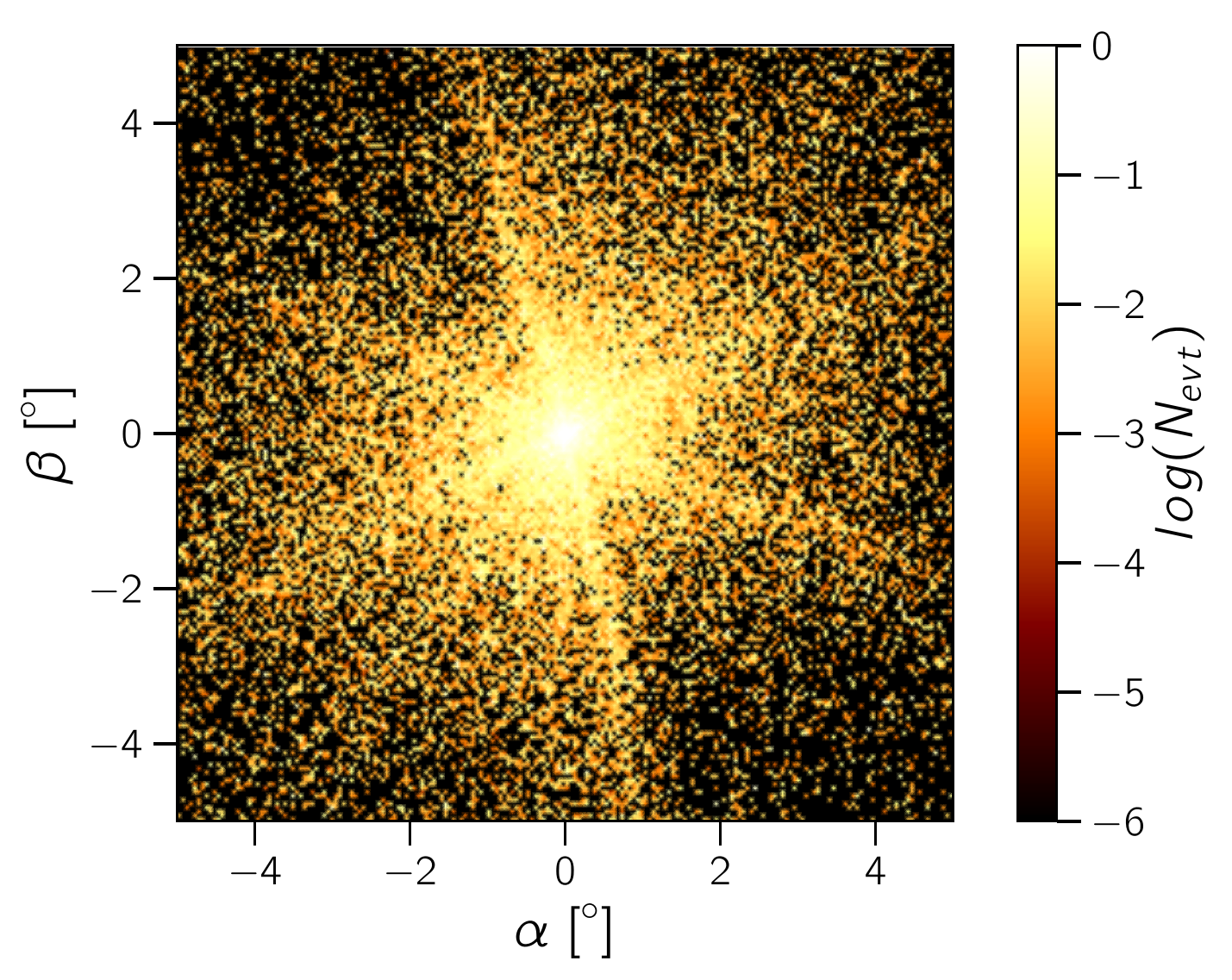}
  \caption{Haloes in the simulation's frame assuming a source at $z \approx 0.14$ emitting 10 TeV gamma rays. The magnetic field is assumed to have a coherence length $L_c = 40 \; \mathrm{Mpc}$, and strength $B=10^{-16} \; \mathrm{G}$, with $\alpha = -3.7$ (left), $\alpha = -3.7$ (middle), and $L_c = 40 \; \mathrm{Mpc}$ (right panel).}
  \label{fig:haloes2}
\end{figure}

While the magnetic field strength plays a role in the shape of the observed haloes, the coherence length determines its shape. In particular, for coherence lengths much larger than the typical mean free path for interactions, the direction of the magnetic field in the site where the first pairs are produced leads to a behaviour similar to that of a uniform magnetic field, since the bulk of the electrons of the cascade would be produced in this region and deflected by the local magnetic field therein. This can be seen in Fig.~\ref{fig:haloes}.
In the right-most panel of Fig.~\ref{fig:haloes} ($L_c = 40 \; \mathrm{Mpc}$) there are clearly preferential directions. As the coherence length decreases, an ellipsoidal shape is still noticeable (central panel, $L_c = 1 \; \mathrm{Mpc}$). For small coherence lengths ($L_c = 30 \; \mathrm{kpc}$) the halo approaches the shape of a circle. The magnetic power spectrum also has an impact in the shape of the morphology of the halo, as shown in Fig.~\ref{fig:haloes2}. Note that these haloes were plotted for the same total number of observed events and the same random seed was used to generate the magnetic fields.

\section{The case of helical magnetic fields}

In this section we briefly address one of the factors that mostly impact the shape of pair haloes: the magnetic helicity. We consider now a source distant 1 Gpc emitting gamma rays of energy $E=10^{13} \;\mathrm{eV}$. Because we are interested in the detectability of helical fields, we now present our results in the observer's frame, i.e., as viewed from Earth.

\begin{figure}[htb!]
  \centering
  \includegraphics[width=0.49\columnwidth]{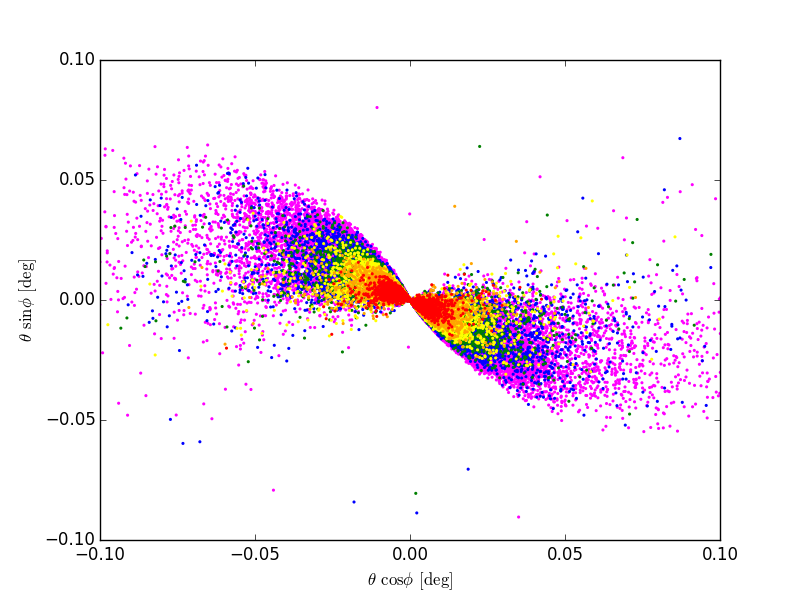}
  \includegraphics[width=0.49\columnwidth]{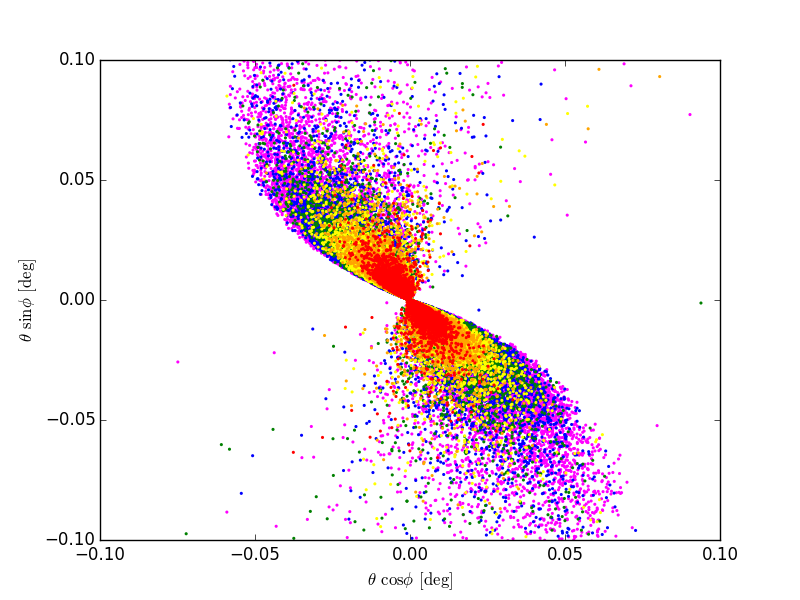}
  \caption{Haloes as observed at Earth (observer's frame) for a source distant 1 Gpc emitting 10 TeV gamma rays. The magnetic field is assumed to have a coherence length $L_c = 225 \; \mathrm{Mpc}$, with spectral index $\alpha = 2$. The left panel corresponds to the case of maximally positive helicity, and the left panel to maximally negative. The colours represent different energy bins: 5-10 GeV (magenta); 10-15 GeV (blue); 15-20 GeV (green); 20-30 GeV (yellow); 30-50 GeV (orange); and 50-100 GeV (red).
  For further details see Ref.~\cite{alvesbatista2016b}.}
  \label{fig:hel1}
\end{figure}

\begin{figure}[htb]
  \centering
  \includegraphics[width=0.49\columnwidth]{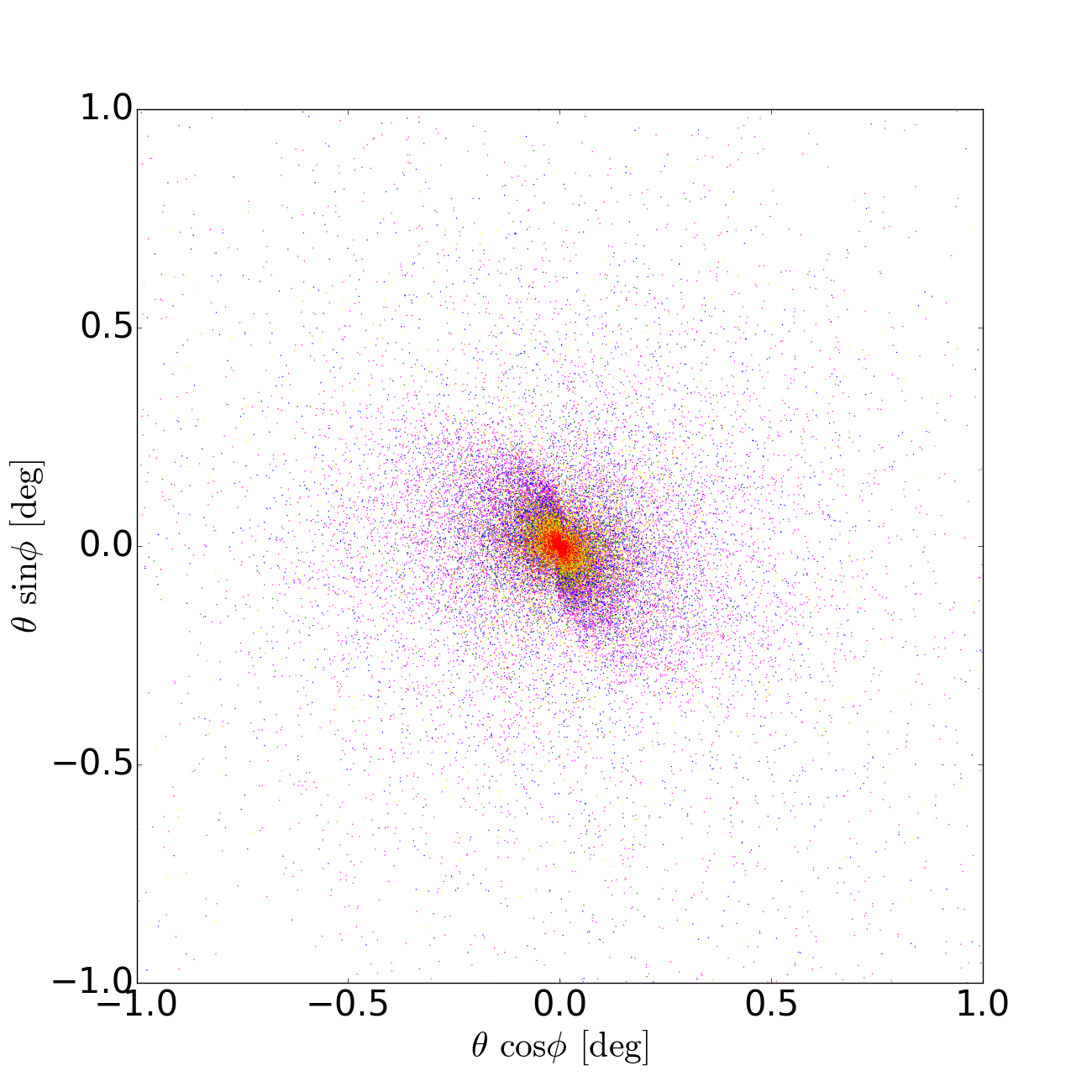}
  \includegraphics[width=0.49\columnwidth]{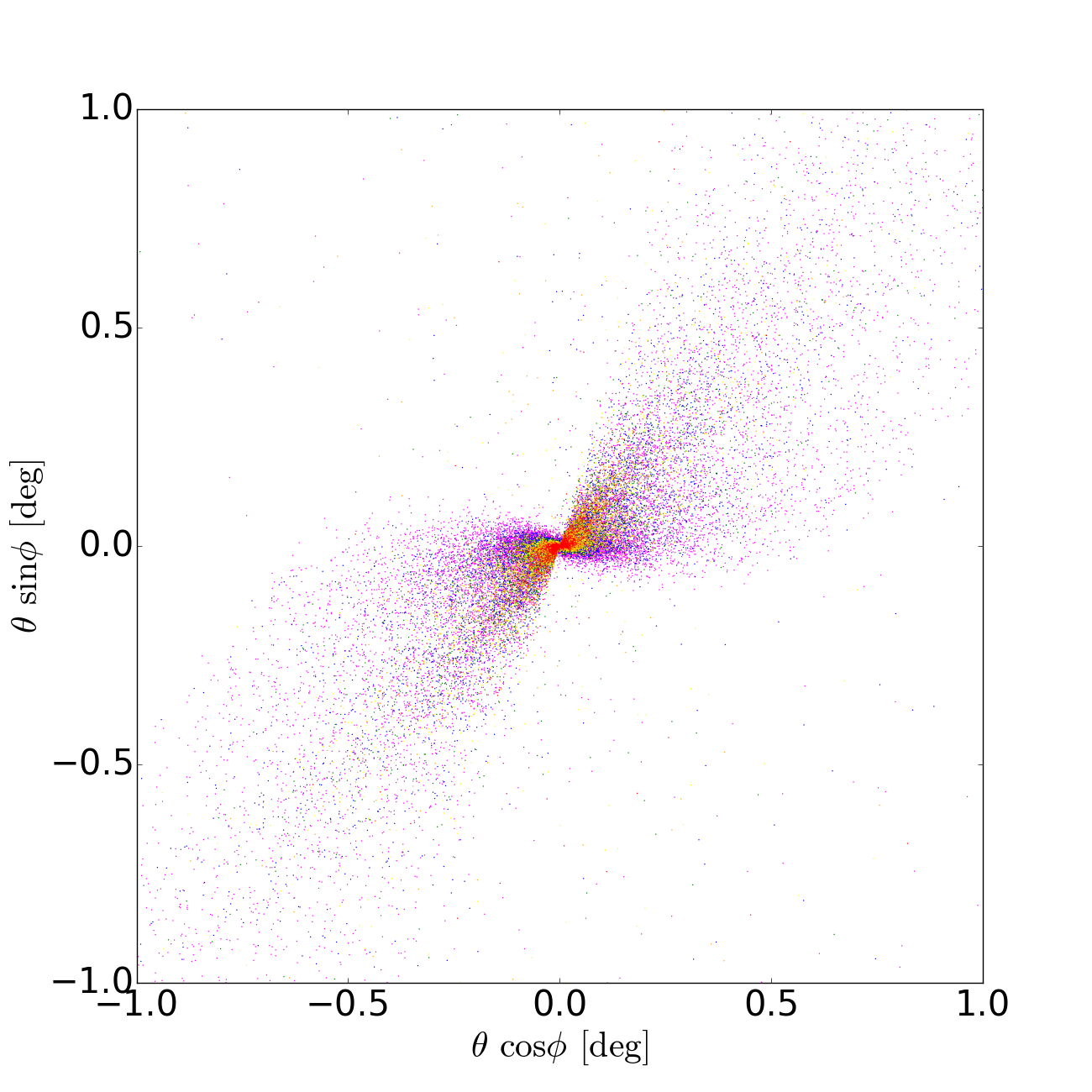}
  \caption{Haloes as observed at Earth (observer's frame) for a source distant 1 Gpc emitting 10 TeV gamma rays. The magnetic field is assumed to have strength $B = 10^{-15} \; \mathrm{G}$, with spectral index $\alpha = 2$. Colours are the same as in Fig.~\ref{fig:hel1}. The left panel corresponds to $L_c = 150 \; \mathrm{Mpc}$, and the left panel to $L_c = 250 \; \mathrm{Mpc}$.
  For further details see Ref.~\cite{alvesbatista2016b}.}
  \label{fig:hel2}
\end{figure}

In Fig.~\ref{fig:hel1} we show how opposite helicities invert the handedness of the halo pattern. We have used a large coherence length ($L_c = 225 \; \mathrm{Mpc}$). 
If this value is decreased, the arrival direction pattern starts to be washed out, completely vanishing for $L_c \lesssim 10 \; \mathrm{Mpc}$, as shown in Fig.~\ref{fig:hel2}.

The spiral-like shapes of the arrival directions shown in Figs.~\ref{fig:hel1} and~\ref{fig:hel2} can be properly quantified using the methods proposed in Refs.~\cite{tashiro2013a,alvesbatista2016b}. 

\section{Summary and outlook}

Using simulations of three-dimensional propagation of high-energy gamma rays, we have considered a few scenarios obtained from the combination of different parameters, namely magnetic field strength, coherence length, and power spectrum. We have verified the suppression of the the flux at $E \lesssim 100 \; \mathrm{GeV}$ for stronger IGMFs. We have also demonstrated that the strength of IGMFs affect the angular size of haloes, whereas the coherence length relates to its shape. Furthermore, we have discussed the role played by the magnetic power spectrum on the formation of haloes. We have not considered the case of misaligned jets, which affects the arrival directions of gamma rays; this will be subject of future works.

We have computed the arrival directions of gamma rays for helical IGMFs. Our results suggest that they have unique signatures in gamma rays if the coherence length is large enough ($L_c \gtrsim 100 \; \mathrm{Mpc}$). Note that while this value may seem rather large, it is not excluded by measurements and most constraints on the coherence length of IGMFs neglect the possible helical nature of the fields. 

Searches for haloes are currently being done using the current generation of IACTs. The upcoming Cherenkov Telescope Array (CTA) offers good prospects for detecting haloes, studying their morphology, and constraining the strength of IGMFs~\cite{cta2017a}.

\section*{Acknowledgements}

The work of A.S. was supported by the Russian Science Foundation under grant no. 17-71-10040, carried out at the Immanuel Kant Baltic Federal University. R.A.B. is supported by a grant from the John Templeton Foundation.

\end{document}